\documentclass[a4paper,UKenglish,cleveref,autoref]{lipics-v2021}

\usepackage{booktabs}

\title{When Review Alone No Longer Scales: Layered Supervision in AI-Assisted Software Engineering}

\titlerunning{Layered Supervision in AI-Assisted Software Engineering}

\author{Markus Stolze}{OST Eastern Switzerland University of Applied Sciences, Rapperswil, Switzerland}{markus.stolze@ost.ch}{https://orcid.org/0000-0003-3506-8024}{}
\author{Mirco Strässle}{smartive AG, St.~Gallen, Switzerland}{mirco@smartive.ch}{}{}

\authorrunning{M.\,Stolze and M.\,Strässle}
\Copyright{Markus Stolze and Mirco Strässle}

\ccsdesc[500]{Software and its engineering~Software development techniques}
\ccsdesc[500]{Software and its engineering~Software verification and validation}
\ccsdesc[300]{Human-centered computing~Empirical studies in HCI}

\keywords{AI-assisted software development, AI coding tools, validation, code review, software teams, guardrails}

\category{Software Engineering in Practice Track Paper}

\relatedversion{}

\supplementdetails[subcategory={Evidence table, codebook, and survey instrument}]{Dataset}{https://doi.org/10.5281/zenodo.21611622}

\acknowledgements{We thank the interview participants for their time and candor, and colleagues at OST who provided feedback on early drafts and the survey instrument.}

\nolinenumbers

\EventEditors{Robert Feldt, Maria Paasivaara, Daniel Mendez, Stefan Wagner, and Marvin Mu\~{n}oz Bar\'{o}n}
\EventNoEds{5}
\EventLongTitle{20th International Symposium on Empirical Software Engineering and Measurement (ESEM 2026)}
\EventShortTitle{ESEM 2026}
\EventAcronym{ESEM}
\EventYear{2026}
\EventDate{October 8--9, 2026}
\EventLocation{Munich, Germany}
\EventLogo{}
\SeriesVolume{394}
\ArticleNo{89}

\begin{document}

\maketitle

\begin{abstract}
AI-assisted development tools enable software engineers to generate implementations at substantially higher speed and volume than in traditional workflows. Software teams have long relied on guardrails --- standing control mechanisms such as code review, linting, testing, and CI/CD pipelines --- to maintain quality and coordination. High-throughput AI-assisted generation increases pressure on these guardrails --- straining their capacity to keep pace with the volume and rate of generated changes --- and reshapes how organizations supervise development workflows, yet relatively little is known about how existing guardrails evolve in response.

We conducted a qualitative interview study with five software engineering practitioners, situated within a broader practitioner survey. Our findings indicate that organizations distribute the work of supervision across multiple guardrail layers: preventive guardrails (produced by externalizing architectural intent and conventions into machine-interpretable form), executable guardrails (linting, testing, and CI/CD repurposed as scalable supervision infrastructure), and human oversight (shifting from line-by-line inspection toward supervisory interpretation focused on architectural reasoning, explainability, and long-term maintainability). We characterize this as a transition from review-centric guardrails toward \emph{layered supervision}, in which no single guardrail carries the supervision load alone.
\end{abstract}

\section{Introduction}

AI-assisted development tools enable software engineers to generate implementations at substantially higher speed and volume than in traditional development workflows. By \emph{AI-assisted development tools} we refer to large language model-based coding assistants and agentic development systems --- for example, IDE-integrated assistants, chat-based code generation tools, and autonomous coding agents --- that generate, modify, or refactor production code based on natural-language or code-context input. Recent advances in such tools have intensified discussions regarding productivity gains, code generation quality, and changing software engineering practices \cite{peng2023impact,zhang2023copilot}. At the same time, AI-assisted generation also raises new challenges regarding how generated artifacts are constrained, validated, and supervised within organizational development environments.

Software engineering workflows have long relied on \emph{guardrails} --- standing control mechanisms such as code review, linting, testing, CI/CD pipelines, and architectural validation --- to support software quality, coordination, and maintainability \cite{jureczko2020codereview,spadini2019tdcr,foalem2026pac}. AI-assisted generation, however, substantially increases implementation throughput while producing artifacts that may remain difficult to validate contextually despite appearing locally plausible or functionally correct \cite{tang2023copilot}. Organizations therefore face growing pressure to supervise AI-assisted development at scales that cannot rely on downstream human review alone. Throughout, we use \emph{guardrail} for any standing control that constrains or checks generation, whether preventive (e.g., specifications, steering files) or detective (e.g., linting, tests, code review), and whether enacted by automated tooling or by humans, and \emph{supervision} for the overarching function of monitoring and steering generation that these guardrails collectively serve. We adopt this term deliberately over narrower alternatives such as \emph{quality assurance mechanism} or \emph{software testing practice}, which do not naturally extend to preventive, upstream constraints such as specifications or steering files that act before any output exists to test.

While prior research has focused strongly on productivity, developer interaction with coding assistants, and validation of generated code \cite{peng2023impact,zhang2023copilot,tang2023copilot}, comparatively few studies examine how organizations operationally constrain and supervise AI-assisted software engineering workflows. In particular, relatively little is known about how existing software engineering guardrails evolve in response to high-throughput AI-assisted implementation activity.

To address this gap, we conducted a qualitative interview study with five software engineering practitioners.

\section{Related Work}
\label{sec:related-work}

\subsection{AI-Assisted Software Engineering}
\label{sec:rw-ai-assisted}

Early empirical work on AI-assisted coding focused on productivity and developer--tool interaction, including controlled studies of GitHub Copilot~\cite{peng2023impact}, qualitative studies of developer validation behavior~\cite{tang2023copilot}, and broader practitioner studies~\cite{zhang2023copilot}. More recent industry-scale evidence indicates that AI-assisted development now produces code at a volume incompatible with exhaustive human inspection: the 2025 DORA report finds approximately 90\% developer adoption and frames AI as a multiplier of existing engineering capability rather than a standalone productivity intervention~\cite{dora2025}, while a large-scale study of 302{,}600 AI-attributed commits across 6{,}299 GitHub repositories reports that AI-introduced quality issues persist and accumulate as technical debt~\cite{liu2026debt}. Studies of agentic coding tools document mixed effects and call explicitly for provenance, supervision, and selective deployment~\cite{agarwal2026agents}, while complementary qualitative work surfaces practitioner concerns about trust and governance in agentic workflows~\cite{vukovic2026enterprise}. These quality-risk findings are not unanimous, however: a preregistered two-phase experiment with 151 professional developers found no significant difference in subsequent maintainability -- measured by independent developers evolving the resulting code -- between AI-assisted and unassisted contributions~\cite{borg2026echoes}, suggesting that downstream quality effects depend on task and context rather than being a uniform property of AI assistance. Our study characterizes how organizations operationally respond to these pressures.

\subsection{Externalized Constraints and Executable Governance}
\label{sec:rw-governance}

Software engineering has long relied on linting, testing, CI/CD pipelines, and architectural validation to support quality and coordination~\cite{jureczko2020codereview,spadini2019tdcr}. A complementary thread examines the externalization of organizational and architectural intent into machine-interpretable form. Foalem et al.\ provide the first large-scale empirical study of Policy-as-Code adoption across 399 open-source repositories~\cite{foalem2026pac}, and practitioner-oriented work on architectural fitness functions and architecture-as-code similarly emphasizes a shift from governance-by-inspection toward governance-by-rule~\cite{ford2026aac}. In AI-assisted development specifically, recent work examines developer-provided context artifacts such as AGENTS.md files, Cursor rules, and structured specification files as configuration substrates for agentic tools~\cite{galster2026configuring} --- the same authors have since retitled a later revision of this study \emph{Harness Engineering for Agentic AI Coding Tools} (arXiv:2602.14690), converging with practitioner discourse that treats a coding agent's harness, everything beyond the model itself, as an attempt to ``externalise and make explicit what human developer experience brings to the table''~\cite{bockeler2026harness}, a formulation that closely echoes our own use of externalization for the preventive and executable guardrail layers below. Osmani situates this harness one layer below what he terms loop engineering, the design of automated, self-triggering agent cycles operating with reduced moment-to-moment human involvement~\cite{osmani2026loop}; our third layer, human oversight, addresses precisely the coordination question this raises as agentic workflows move toward such loops. Our study examines how such externalized constraints proactively guide AI-assisted generation upstream of review.

\subsection{Human Oversight of AI-Assisted Development}
\label{sec:rw-oversight}

A growing literature examines developer trust, automation bias, and oversight behavior in AI-assisted workflows. Qualitative studies report that developers often over-trust AI suggestions and may reduce vigilance when code appears plausible, particularly with respect to security~\cite{klemmer2024aiassistants}. Broader work on human oversight of AI systems, including Article~14 of the EU AI Act, distinguishes monitoring, interpretation, and intervention as separable oversight functions~\cite{fink2025oversight}. Our findings contribute to this discussion by characterizing how oversight in AI-assisted software engineering shifts from post-hoc inspection toward supervisory interpretation focused on architectural reasoning, contextual appropriateness, and long-term maintainability.

\section{Methodology}
\label{sec:method}

\subsection{Study Design and Data Collection}
\label{sec:study-design}

This study employed an exploratory qualitative interview approach to investigate how software engineering practitioners concretely constrain, supervise, and govern AI-assisted development workflows. Semi-structured interviews focused on AI-assisted coding workflows, code review practices, externalization, executable guardrails, and human oversight.

\subsection{Participants}
\label{sec:participants}

We conducted semi-structured interviews with five practitioners selected from a broader practitioner survey (Section~\ref{sec:survey-context}), intentionally including a mix of roles (engineers, architects, technical leads, engineering managers) and organizational contexts (small product companies, large enterprises, agencies, industrial firms). Table~\ref{tab:participants} summarizes the participants; we refer to them as~P1--P5 throughout. The interviews were intended to provide in-depth insight into emerging supervision practices rather than statistically representative coverage.

\begin{table*}[t]
\centering
\caption{Interview participants and organizational contexts. Team size
and governance posture are reported from each participant's survey
response.}
\label{tab:participants}
\small
\begin{tabular}{@{}p{0.06\textwidth}p{0.20\textwidth}p{0.19\textwidth}p{0.13\textwidth}p{0.23\textwidth}@{}}
\toprule
\textbf{ID} & \textbf{Role} & \textbf{Domain} & \textbf{Team size} & \textbf{Governance posture} \\
\midrule
P1 & Software Architect / CTO & Enterprise software & Small (2--9) & Informal, AI-recommended \\
P2 & Staff Engineering Manager & Energy utility frontend & Large (10+) & Clear guidelines \\
P3 & Engineering Team Lead & Construction software & Large (10+) & Clear guidelines \\
P4 & Senior Frontend Engineer & Digital agency & Small (2--9) & Informal arrangements \\
P5 & Senior UI Engineer & Industrial technology & Small (2--9) & No explicit rules \\
\bottomrule
\end{tabular}
\end{table*}

\subsection{Survey Context}
\label{sec:survey-context}

To situate the interview findings, we draw on aggregate responses from a survey conducted in November--December~2025 within our institution's computer science alumni network. The survey was co-developed by the first author and a colleague based on practical experience with AI-assisted tools, reviewed by one external practitioner for clarity, but not formally piloted. Approximately 100 alumni were personally invited; 50 completed the survey. Respondents were predominantly senior software engineers (16/50) and technical leads or architects (26/50), working mostly in small teams of 2--9 (34/50). Because the survey relied on convenience sampling and an unpiloted instrument, we do not interpret its aggregate responses as inferential evidence; we use them to provide indicative context for the qualitative findings.

\subsection{Data Analysis}
\label{sec:data-analysis}

The interview material was analyzed using an iterative thematic analysis approach following the six-phase framework of Braun and Clarke~\cite{braunclarke2006}. Analysis began with close reading of interview transcripts and follow-up email responses, followed by open coding focused on recurring practices, tensions, supervision mechanisms, and organizational responses. Codes were iteratively refined into a scheme of six top-level categories (reconfiguration of validation, transformation of code review, understanding and explainability, guardrails and AI control mechanisms, context and maturity, and risks and failure modes), with three to eleven subcodes each; the full scheme is available in the supplementary material. Findings were continuously mapped back to supporting interview excerpts to strengthen traceability.

\paragraph*{AI-supported analysis.}
The analysis was conducted by a single researcher (the first author), with ChatGPT used under researcher direction for auxiliary tasks: translation of German excerpts into English analytical paraphrases, surfacing recurring phrasing to support code consolidation, and drafting preliminary thematic summaries. All coding decisions, theme construction, evidence-to-theme mapping, and quote selection were made and verified by the researcher against the original transcripts. For the camera-ready revision, Claude (Anthropic) was used under the same direction to re-verify all directly quoted material against the original transcripts and correct several participant-attribution errors this process identified; all resulting changes were reviewed and approved by the first author.

\paragraph*{Member-checking session.}
After the initial round of coding and thematic synthesis, we conducted a structured member-checking session with four of the five interview participants (P1, P2, P4, P5; P3 was unavailable) in late March~2026, in which preliminary findings were presented and discussed. Participants confirmed, qualified, or contradicted specific claims, and contributed additional examples and clarifications. The session was audio-recorded and transcribed with participant consent. Material from the session was incorporated into the final analysis: in particular, several illustrative examples and reframings reported in Section~\ref{sec:findings} and Section~\ref{sec:disc-tradeoffs} originate from the member-checking session rather than from the original interviews.

\paragraph*{Limitations of the analytical approach.}
Single-coder qualitative analysis without independent inter-rater reliability assessment introduces interpretive risk, which we partly mitigated through iterative refinement, written follow-up clarifications with each participant, and the member-checking session described above. We further discuss the limitations of the study, including sample size and the situated nature of the findings, in Section~\ref{sec:limitations}.

\section{Findings}
\label{sec:findings}

We report findings as five themes (F1--F5). Each quoted phrase is attributed to its participant; ``participants'' is reserved for statements supported by all five.

\subsection{F1: Review-centric guardrails no longer scale, and specification quality becomes consequential}
\label{sec:f1}

Participants consistently described AI-assisted development as substantially accelerating implementation while leaving review and validation constrained by human contextual reasoning. As P2 emphasized, code writing has become \textit{``very cheap''}~[P2] under AI-assisted development, with \textit{``the bottleneck clearly moving to review''}~[P2].
P1 quantified this concretely: manual review of an AI-assisted one-shot implementation typically takes \textit{``three to four times as long''}~[P1] as the generation itself (Section~\ref{sec:f2}) --- a ratio between the review and generation phases of the AI-assisted workflow, not a before/after-AI comparison.
P1, P2, P4, and~P5 described AI-assisted development as intensifying rather than reducing pressure on review-centric practices. For context, code quality and long-term maintainability were the most frequently selected risks in the survey (43 and~40 of 50~respondents respectively).

P1 observed that what has changed is not the existence of upstream specification work, which experienced teams have long invested in~\cite{ford2026aac}, but its \emph{consequentiality}: specification quality now propagates directly into generated code, with weak input producing weak output that LLMs do not push back against the way an experienced engineer would. This makes upstream effort more consequential, not merely more abundant.

A recurring concern involved generated solutions that appeared correct at first glance while remaining difficult to validate contextually --- a phenomenon we refer to as \emph{false correctness}. P2, P4, and~P5 described generated artifacts as syntactically correct and locally coherent, yet problematic regarding architectural consistency, explainability, or long-term maintainability. P2 reported being unable to change parts of a system because he could no longer trace \textit{``why they are now where they are''}~[P2], and P5 summarized the pattern in words he had already used in his own written survey response before the interview: \textit{``functionally correct, but nobody understands why''}~[P5]. False correctness shifted the burden of validation away from behavioral testing toward reasoning about architectural intent, contextual consistency, explainability, and long-term maintainability.

\subsection{F2: Teams externalize implicit context into upstream guardrails}
\label{sec:f2}

P4 and~P5 described existing repositories, documentation, and architectural guidelines as often insufficient for AI-assisted generation: AI systems inferred incomplete, outdated, or locally plausible patterns without reliably capturing intended architectural direction. P5 further noted that different artifacts and implementation signals could contradict each other or evolve independently, creating ambiguity about authoritative guidance.

In response, participants described externalizing contextual intent more explicitly before generation. P1 estimated that \textit{``two thirds of the time go into specification refinement''}~[P1], and P4 confirmed a similar shift in his own team: substantially longer architectural planning at the start of a task, with correspondingly shorter time spent actually writing code~[P4].
These practices also appeared selectively in the survey: 13/50 respondents reported using steering files such as \texttt{.cursor/config} or \texttt{claude.md}, and 6/50 reported structured prompt workflows (Requirements~$\rightarrow$~Design~$\rightarrow$~Tasks)~--- suggesting that externalization practices are emerging but not yet widespread.

P4 emphasized that once a recurring constraint was identified, it had to be handed off to automatic enforcement: \textit{``If a rule is relevant, it must be enforced through linting''}~[P4] --- the point at which externalization (identifying and formalizing the constraint) meets executable guardrails (Section~\ref{sec:f3}).
P1 additionally described orchestration sub-agents that monitor for drift during generation (discussed further in Section~\ref{sec:implications}). These practices convert tacit knowledge into reusable \emph{preventive} guardrails that shape generation upstream, rather than only \emph{detective} guardrails that catch problems downstream. P4 and~P5 noted important limits, however: not all architectural assumptions or organizational conventions can be formalized in advance, which is why teams complemented these upstream guardrails with executable ones embedded in the build (Section~\ref{sec:f3}).

\subsection{F3: Executable guardrails expand in importance}
\label{sec:f3}

P1, P2, P4, and~P5 described linting, unit testing, CI/CD pipelines, and automated validation as increasingly critical guardrails for AI-assisted workflows. P1 and~P2 reported extending existing executable enforcement with additional checks detecting architectural violations, code duplication, dependency misuse, and deviations from organizational implementation constraints. P2 and~P4 described how such executable enforcement made the build system itself the arbiter: violations of encoded rules would break the build rather than being caught by human reviewers. P5 illustrated the deliberate character of this stance in his own monorepo: he consciously chose not to relax automated conventions even though doing so would have been easy --- \textit{``as long as I can do it automatically it costs me nothing\,\ldots\,I really want this convention to be strictly upheld''}~[P5]. These executable guardrails were not limited to syntactic correctness but extended toward architectural conformance and organizational implementation policies.

P4 emphasized important limits: contextual interpretation, architectural tradeoff reasoning, and long-term maintainability assessment frequently could not be checked automatically. Executable guardrails thus carried part of the supervision load but did not replace human oversight, reducing recurring review burden while leaving contextual and interpretive assessment unresolved.

\subsection{F4: Human oversight shifts toward supervisory and operational explainability}
\label{sec:f4}

Despite increasing reliance on preventive and executable guardrails, participants consistently rejected fully autonomous workflows. P1, P2, and~P4 emphasized that AI-generated artifacts continued to require a human guardrail layer in situations involving architectural tradeoffs, contextual ambiguity, or organizational constraints that could not be encoded reliably. As P1 summarized, AI-assisted development was \textit{``no fire-and-forget''}~[P1].

P1 described the human guardrail as more supervisory and intervention-oriented: rather than reviewing artifacts only after implementation completion, developers \textit{``observe step-by-step\,\ldots\,interrupt early''}~[P1] when generation diverged from intended direction. Here intervention is an act of supervision triggered when a guardrail --- whether an automated drift-detector or the observing developer --- surfaces a problem. P4 additionally emphasized that the human guardrail focused on architectural reasoning and long-term maintainability rather than complete line-by-line comprehension. In place of full comprehension, P4 described \emph{operational explainability}: the requirement that generated systems remain diagnosable and reconstructable on demand, supported by abstraction layers, visualization mechanisms, logs, and higher-level interpretive tooling that make the human guardrail effective without line-by-line reading.

The strength of the explainability requirement varied by stakes. P1 noted that organizational tolerance differed sharply by use case: a short-lived UX prototype may not require any line-level developer understanding, whereas a database migration or a feature in a system with strict recovery-time requirements demands precise comprehension of every statement. Operational explainability is therefore not a single threshold but a context-dependent capability calibrated to system criticality and longevity.

P2 and~P4 raised concerns about the sustainability of such interpretive oversight, particularly for less experienced developers, a concern recently elevated to a broader profession-level question by Russinovich and Hanselman~\cite{russinovich2026cacm} (see also Section~\ref{sec:limitations}). Survey responses indicated related concerns: 23/50 respondents wished for targeted support on review standards for AI-generated code, and 15/50 on monitoring and traceability tools.

\subsection{F5: Guardrail configurations vary along identifiable dimensions}
\label{sec:f5}

The five participants varied substantially in how they combined the three guardrail layers --- preventive guardrails, executable guardrails, and human oversight. We identify three dimensions along which configurations differed: (i)~\emph{governance posture} --- the presence and formality of organizational AI-tool guidelines; (ii)~\emph{system criticality and homogeneity} --- the longevity and consistency of the codebase under AI-assisted modification; and (iii)~\emph{team composition} --- the distribution of senior and less experienced developers and the resulting supervisory capacity. We deliberately avoid the term ``maturity'' here, since it suggests a linear progression that our data does not support.

Along the governance dimension, P2 and~P3 described environments with clear, formal AI-tool guidelines and structured review procedures, whereas P1, P4, and~P5 reported predominantly informal arrangements or no explicit rules. Survey responses suggest similar heterogeneity beyond the interview sample: asked how AI-tool use is governed or monitored, 24 of the 50 respondents reported clear organizational guidelines and 23 reported only informal arrangements or no monitoring at all. Asked separately how collaboration had changed, 8~respondents reported that responsibility for AI-assisted code had become more diffuse.

System criticality and homogeneity strongly shaped which mechanisms worked. P1 explicitly framed his team's practices as contingent on a modern, low-legacy, highly conformant codebase, and was explicit that the same mechanisms would likely not transfer to grown legacy systems. P5 correspondingly described legacy repositories as particularly challenging because inconsistent patterns and undocumented assumptions increased ambiguity.

P2 and~P5 observed that teams with stronger architectural expertise used executable constraints more extensively. The lower bound is worth naming: teams that adopt AI tools widely without proportionate senior supervisory capacity may accumulate risk not visible from within the team, a concern consistent with broader observations on AI's seniority-biased effects~\cite{russinovich2026cacm}.

\section{Discussion}
\label{sec:discussion}

\subsection{From Review-Centric to Layered Supervision}
\label{sec:disc-layered}

Participants responded to the pressures identified in Section~\ref{sec:findings} by distributing supervision across three guardrail layers: preventive guardrails, executable guardrails, and human oversight. No single layer proved sufficient: preventive guardrails cannot capture all architectural tradeoffs in advance, executable guardrails cannot evaluate contextual appropriateness, and the human guardrail cannot scale to AI-generation throughput.

\subsection{Pressure Toward Upstream Externalization}
\label{sec:disc-operationalization}

Traditional artifacts such as repositories, documentation, and architectural guidelines relied on human interpretation to be useful; in AI-assisted workflows, we read our data as indicating that this interpretive layer is largely absent in the traditional sense --- not because AI systems process context without any interpretive mechanism at all (a probabilistic model plausibly performs a form of interpretation of its own), but because the human sense-making step that traditionally connected a written artifact to a considered action is no longer a reliable intermediary. Specification refinement, architectural planning, orchestration workflows, and the preventive guardrails they produce therefore take on a new role: they shape permissible generation trajectories ahead of implementation rather than only documenting decisions already made. Externalization does not eliminate the need for supervision, however: contextual assumptions, architectural tradeoffs, and organizational expectations can only ever be partially formalized, and it is often unclear in advance whether a given level of externalization is sufficient to reliably steer AI-assisted generation.

A natural question is where externalization ends and executable guardrails (Section~\ref{sec:f3}) begin, since both act on generation before a human reviews the output. We distinguish the two by mechanism rather than by sequence. Externalization produces preventive guardrails --- specifications, steering files, architectural plans --- that guide what an AI system generates but are not themselves automatically checked; their effect depends on whether the generation process actually consults them. Executable guardrails, by contrast, are automatically evaluated against generated output regardless of how that output was produced. In practice the two run concurrently rather than as sequential stages: a steering file (a preventive guardrail produced through externalization) encoding a convention and a linting rule enforcing the same convention (an executable guardrail) typically coexist, so that the executable check still catches violations when the steering artifact is outdated, ignored, or absent from a particular generation trajectory. Externalization thus shapes generation \emph{ex ante}; executable guardrails verify it \emph{ex post}, independent of whether the ex-ante shaping succeeded.

\subsection{What Changes Relative to Pre-LLM Software Engineering Governance}
\label{sec:disc-whats-new}

The individual mechanisms participants described --- linting,
CI/CD-based enforcement, architectural fitness functions, policy-as-code,
specification-driven workflows --- are not in themselves new. Each has
a substantial pre-LLM history in software engineering practice and
research~\cite{jureczko2020codereview,spadini2019tdcr,foalem2026pac,ford2026aac}.
Our findings suggest that what is changing is not the existence of these
mechanisms, but their operational role, relative weight, and
configuration.

We highlight three shifts. Guardrails move \emph{upstream}: steering files, structured prompts, and specification refinement increasingly act as \emph{preventive} guardrails that guide generation, rather than the field's traditional reliance on \emph{detective} guardrails that inspect finished code (cf.\ Section~\ref{sec:f2}). Executable guardrails are \emph{promoted in priority}: P4 articulated a stance in which any sufficiently relevant rule is treated as a candidate for automatic linting enforcement (Section~\ref{sec:f2}), reserving the human guardrail for what cannot be encoded (Section~\ref{sec:f3}). Human oversight is \emph{redistributed temporally}: P1 described monitoring generation step by step and intervening early~[P1], shifting the human guardrail from post-hoc inspection toward concurrent supervision (Section~\ref{sec:f4}). These shifts amplify and reconfigure existing practices under AI-induced throughput and contextual-opacity pressures, rather than replacing them.

\subsection{Human Oversight Shifts from Inspection to Supervisory Interpretation}
\label{sec:disc-oversight}

The reframing of oversight described in Section~\ref{sec:f4} carries an implication for expertise. Effective supervisory interpretation, abstraction-based assessment, and operational explainability all rest on architectural and contextual judgment that participants framed as in short supply. AI-assisted development may therefore shift professional expertise toward architectural reasoning, contextual interpretation, and supervisory intervention, with corresponding implications for how less experienced developers acquire this expertise --- a profession-level question we return to in Section~\ref{sec:implications}.

\subsection{Risk--Productivity Trade-offs in Practice}
\label{sec:disc-tradeoffs}

The adoption of AI-assisted development introduces a tension between productivity gains and emerging validation costs. Participants described faster implementation and reduced effort in code-writing phases as genuine and substantial. As P4 emphasized during the member-checking session, organizations would not currently be investing this level of attention if the productivity gains were marginal --- an inference we draw from participants' observed investment of time and organizational effort rather than a claim any participant stated directly in these terms.
At the same time, these gains were highly contingent on guardrails. Without sufficient preventive, executable, and human guardrails, the gains were frequently offset by increased review effort, accumulated false correctness, and delayed defect detection.

P2 recounted one such trajectory: in a project where AI-generated changes accumulated over several months without proportionate review capacity, a substantive feature had to be discarded and reimplemented from scratch once its architectural problems became visible. P2 framed this as a failure mode rather than an inherent property of AI-assisted development: \textit{``had this been done without an AI system, we would not have generated so much code\,\ldots\,maybe we would have noticed earlier''}~[P2]. The lesson is not that AI-assisted generation is unsafe, but that productivity gains and validation infrastructure need to be scaled together --- an organizational choice rather than a purely technical optimization.

\subsection{Implications for Software Engineering Practice}
\label{sec:implications}

The participants' practices suggest several concrete patterns rather than recommendations, since they emerged in specific organizational contexts and require adaptation rather than direct transfer.

\paragraph*{Pattern 1: Promote recurring review findings into linting rules.}
P4 articulated a stance in which review effort is reserved for what cannot be encoded automatically (quoted in full in Section~\ref{sec:f2}); P5 exemplified the operational consequence with the same deliberate strictness: \textit{``I could leave it out --- but I consciously chose not to. I really want this convention to be strictly upheld''}~[P5]. The pattern is to treat each manually-recurring review concern as a candidate for promotion into an executable check, narrowing manual review toward the genuinely contextual. This aligns with established policy-as-code~\cite{foalem2026pac} and architectural fitness-function practice~\cite{ford2026aac}, applied as an explicit response to AI-generation throughput.

\paragraph*{Pattern 2: Front-load specification and architectural planning.}
P1 estimated that \textit{``two thirds of the time go into specification refinement''}~[P1], with P4 reporting longer architectural planning at the start of a task and considerably shorter coding time~[P4]. Effort that previously sat in implementation has migrated upstream; organizations should budget for this redistribution rather than treating planning as overhead to compress.

\paragraph*{Pattern 3: Use sub-agents for concurrent supervision and independent review.}
Two complementary mechanisms emerged in P1's team. First, planner sub-agents constrain generation upstream: P1 described using \textit{``three different planner sub-agents''}~[P1] whose proposals are ranked by a further agent, with the explicit purpose of detecting when \textit{``an agent has completely drifted''}~[P1] from the intended plan -- decoupling architectural-conformance checking from post-hoc review by embedding it concurrently with generation.

\paragraph*{Pattern 4: Shift validation toward higher-level abstractions.}
In some teams, the primary response to AI-generated code volume is not deeper line-by-line inspection but \emph{validation through higher-level artifacts} --- architecture plans, visual system overviews, sequence diagrams, and other intermediate representations that make large changes easier to reason about and verify. P4 framed this as particularly important in agency contexts with heterogeneous projects and limited review capacity, where teams shift effort from code-level toward architecture-level validation, sometimes using AI-generated visual overviews to check intended structure, change scope, and system fit during implementation. Returns on tooling that makes architectural state and implementation trajectories \emph{legible at a glance} appear larger than returns on tooling that merely accelerates generation.

\paragraph*{A note on expertise.}
Across all patterns, effective application appeared to depend on architectural and contextual expertise that participants framed as in short supply (P2, P4). Whether organizations can sustain this expertise as implementation activities become increasingly automated is a broader question beyond the scope of this paper; we refer interested readers to recent profession-level discussions of the issue~\cite{russinovich2026cacm}.

\subsection{Limitations and Future Work}
\label{sec:limitations}

The findings are based on five interviews drawn from a single practitioner survey conducted within an alumni network through convenience sampling and an unpiloted instrument (Section~\ref{sec:survey-context}); the practitioner population over-represents Swiss and Central European contexts, and practices in this space evolve rapidly. The analysis was conducted by a single researcher with AI-supported auxiliary work; we mitigated single-coder risk through iterative refinement, follow-up clarifications, and the member-checking session described in Section~\ref{sec:data-analysis}, though no independent inter-rater reliability assessment was performed.

One of the five interview participants (P4) is also a co-author of this paper, contributing the practitioner perspective required by the SEIP track. This dual role creates two distinct risks of bias. First, P4's own statements could have been selected or framed more favorably than other participants' during quote selection and writing. Second, P4's involvement in the project could have reshaped the coding scheme or thematic framing around his own contributions rather than the full dataset. We mitigated the first risk by having the first author verify all quotations attributed to P4 against the original interview and member-checking transcripts independently, without P4's input on selection or wording. We mitigated the second by finalizing the coding scheme and thematic framing before P4 joined as co-author, so that his subsequent contribution was limited to reviewing and co-writing the manuscript rather than to re-deriving the analysis.

The study captures reported practices and perceptions rather than direct longitudinal observation. Future work could complement interview-based findings with ethnographic observations, repository analyses (cf.~\cite{liu2026debt,agarwal2026agents}), or longitudinal studies; cost-economic drivers of AI adoption, identified by participants as primary motivations but outside our analytic scope, also warrant investigation. Due to confidentiality agreements, full interview transcripts cannot be shared; the coding scheme, survey instrument, and anonymized evidence table are openly available in our supplementary Zenodo repository (\url{https://doi.org/10.5281/zenodo.21611622}).

\section{Conclusion}

AI-assisted development appears not to automate implementation in isolation but to redistribute supervision across the development process. Based on a qualitative interview study with five practitioners and a member-checking session, we identified a transition from review-centric guardrails toward \emph{layered supervision}, in which the supervision function is distributed across three guardrail layers: preventive guardrails, executable guardrails, and human oversight. Existing engineering infrastructure --- linting, testing, CI/CD, architectural validation --- gains a new role as scalable supervision infrastructure for AI-assisted generation, while the human guardrail shifts from line-by-line inspection toward supervisory interpretation focused on architectural reasoning, explainability, and long-term maintainability.

\bibliography{LIPIcs-ESEM2026-89}

@article{peng2023impact,
  title={{The Impact of AI on Developer Productivity: Evidence from GitHub Copilot}},
  author={Peng, Sida and Kalliamvakou, Eirini and Cihon, Peter and Demirer, Mert},
  journal={arXiv preprint arXiv:2302.06590},
  year={2023},
  doi={10.48550/arXiv.2302.06590},
  url={https://doi.org/10.48550/arXiv.2302.06590}
}

@inproceedings{zhang2023copilot,
  title={{Practices and Challenges of Using GitHub Copilot: An Empirical Study}},
  author={Zhang, Beiqi and Liang, Peng and Zhou, Xiyu and Ahmad, Aakash and Waseem, Muhammad},
  booktitle={Proceedings of the 35th International Conference on Software Engineering and Knowledge Engineering (SEKE)},
  pages={124--129},
  year={2023},
  note={arXiv:2303.08733}
}

@inproceedings{tang2023copilot,
  title={{An Empirical Study of Developer Behaviors for Validating and Repairing AI-Generated Code}},
  author={Tang, Ningzhi and Chen, Meng and Ning, Zheng and Bansal, Aakash and Huang, Yu and McMillan, Collin and Li, Toby Jia-Jun},
  booktitle={Proceedings of the 13th Annual Workshop on the Intersection of HCI and PL (PLATEAU)},
  year={2023}
}

@article{jureczko2020codereview,
  title={{Code review effectiveness: an empirical study on selected factors influence}},
  author={Jureczko, Marian and Kajda, {\L}ukasz and G{\'o}recki, Pawe{\l}},
  journal={IET Software},
  volume={14},
  number={7},
  pages={794--805},
  year={2020},
  doi={10.1049/iet-sen.2020.0134},
  url={https://doi.org/10.1049/iet-sen.2020.0134}
}

@inproceedings{spadini2019tdcr,
  title={{Test-Driven Code Review: An Empirical Study}},
  author={Spadini, Davide and Palomba, Fabio and Baum, Tobias and Hanenberg, Stefan and Bruntink, Magiel and Bacchelli, Alberto},
  booktitle={Proceedings of the 41st International Conference on Software Engineering (ICSE)},
  pages={1061--1072},
  year={2019},
  doi={10.1109/ICSE.2019.00110},
  url={https://doi.org/10.1109/ICSE.2019.00110}
}

@article{foalem2026pac,
  title={{An Empirical Study of Policy-as-Code Adoption in Open-Source Software Projects}},
  author={Foalem, Patrick Loic and Khomh, Foutse and Da Silva, Leuson and Merlo, Ettore},
  journal={arXiv preprint arXiv:2601.05555},
  year={2026},
  doi={10.48550/arXiv.2601.05555},
  url={https://doi.org/10.48550/arXiv.2601.05555}
}

@misc{dora2025,
  title={{2025 State of AI-Assisted Software Development}},
  author={{DORA, Google Cloud}},
  year={2025},
  howpublished={\url{https://services.google.com/fh/files/misc/2025_state_of_ai_assisted_software_development.pdf}},
  note={DevOps Research and Assessment (DORA) Report}
}

@article{liu2026debt,
  title={{Debt Behind the AI Boom: A Large-Scale Empirical Study of AI-Generated Code in the Wild}},
  author={Liu, Yue and Widyasari, Ratnadira and Zhao, Yanjie and Irsan, Ivana Clairine and Chen, Junkai and Lo, David},
  journal={arXiv preprint arXiv:2603.28592},
  year={2026},
  doi={10.48550/arXiv.2603.28592},
  url={https://doi.org/10.48550/arXiv.2603.28592}
}

@inproceedings{agarwal2026agents,
  title={{AI IDEs or Autonomous Agents? Measuring the Impact of Coding Agents on Software Development}},
  author={Agarwal, Shyam and He, Hao and Vasilescu, Bogdan},
  booktitle={Proceedings of the 23rd International Conference on Mining Software Repositories (MSR)},
  year={2026},
  note={arXiv:2601.13597},
  doi={10.48550/arXiv.2601.13597},
  url={https://doi.org/10.48550/arXiv.2601.13597}
}

@inproceedings{vukovic2026enterprise,
  title={{Usage, Effects and Requirements for AI Coding Assistants in the Enterprise: An Empirical Study}},
  author={Vukovic, Maja and Pan, Rangeet and Ho, Tin Kam and Krishna, Rahul and Pavuluri, Raju and Merler, Michele},
  booktitle={Proceedings of the 3rd International Workshop on Large Language Models for Code (LLM4Code), co-located with ICSE 2026},
  year={2026},
  doi={10.1145/3786181.3788727},
  url={https://doi.org/10.1145/3786181.3788727}
}

@inproceedings{galster2026configuring,
  title={{Configuring Agentic AI Coding Tools: An Exploratory Study}},
  author={Galster, Matthias and Mohsenimofidi, Seyedmoein and Lulla, Jai Lal and Abubakar, Muhammad Auwal and Treude, Christoph and Baltes, Sebastian},
  booktitle={Proceedings of the 3rd ACM International Conference on AI-Powered Software (AIware)},
  year={2026},
  doi={10.1145/3805760.3814887},
  url={https://doi.org/10.1145/3805760.3814887}
}

@book{ford2026aac,
  title={{Architecture as Code: Quantifying the Intersections of Software Architecture}},
  author={Ford, Neal and Richards, Mark},
  publisher={O'Reilly Media},
  year={2027},
  isbn={9798341640368},
  note={Early Release}
}

@inproceedings{klemmer2024aiassistants,
  title={{Using AI Assistants in Software Development: A Qualitative Study on Security Practices and Concerns}},
  author={Klemmer, Jan H. and Horstmann, Stefan Albert and Patnaik, Nikhil and Ludden, Cordelia and Burton Jr., Cordell and Powers, Carson and Massacci, Fabio and Rahman, Akond and Votipka, Daniel and Lipford, Heather Richter and Rashid, Awais and Naiakshina, Alena and Fahl, Sascha},
  booktitle={Proceedings of the 2024 on ACM SIGSAC Conference on Computer and Communications Security (CCS '24)},
  pages={2726--2740},
  year={2024},
  doi={10.1145/3658644.3690283},
  url={https://doi.org/10.1145/3658644.3690283}
}

@article{fink2025oversight,
  title={{Human Oversight under Article 14 of the EU AI Act}},
  author={Fink, Melanie},
  journal={SSRN Electronic Journal},
  year={2025},
  note={Forthcoming in: \emph{The AI Act Commentary: A Thematic Analysis}, Hart-Bloomsbury, 2026},
  doi={10.2139/ssrn.5147196},
  url={https://doi.org/10.2139/ssrn.5147196}
}

@article{braunclarke2006,
  title={{Using thematic analysis in psychology}},
  author={Braun, Virginia and Clarke, Victoria},
  journal={Qualitative Research in Psychology},
  volume={3},
  number={2},
  pages={77--101},
  year={2006},
  doi={10.1191/1478088706qp063oa},
  url={https://doi.org/10.1191/1478088706qp063oa}
}

@article{russinovich2026cacm,
  title={{Redefining the Software Engineering Profession for AI}},
  author={Russinovich, Mark and Hanselman, Scott},
  journal={Communications of the ACM},
  volume={69},
  number={4},
  year={2026},
  doi={10.1145/3779312},
  url={https://doi.org/10.1145/3779312},
  note={Opinion column}
}

@misc{bockeler2026harness,
  title={{Harness Engineering for Coding Agent Users}},
  author={B{\"o}ckeler, Birgitta},
  howpublished={martinfowler.com},
  year={2026},
  url={https://martinfowler.com/articles/harness-engineering.html},
  note={Part of Martin Fowler's ``Exploring Generative AI'' series; published 2 April 2026, accessed 30 July 2026}
}

@misc{osmani2026loop,
  title={{Loop Engineering}},
  author={Osmani, Addy},
  howpublished={addyosmani.com},
  year={2026},
  url={https://addyosmani.com/blog/loop-engineering/},
  note={Published 7 June 2026, accessed 30 July 2026}
}

@article{borg2026echoes,
  title={{Echoes of AI: Investigating the downstream effects of AI assistants on software maintainability}},
  author={Borg, Markus and Hewett, Dave and Hagatulah, Nadim and Couderc, Noric and S{\"o}derberg, Emma and Graham, Donald and Kini, Uttam and Farley, Dave},
  journal={Empirical Software Engineering},
  volume={31},
  note={Article number 161},
  year={2026},
  doi={10.1007/s10664-026-10889-1},
  url={https://doi.org/10.1007/s10664-026-10889-1}
}

\end{document}